\pdfoutput=1 
\documentclass{JINST}

\newcommand{\etal}{{\it et al.}}
\newcommand{\nancay}{Nan\c{c}ay}

\newcommand{\TorchiSA}{Torchinsky,~S.A.}
\newcommand{\enancay}{\mbox{EMBRACE@\nancay}}
\newcommand{\degrees}{$^\circ$}

\title{\enancay: An Ultra Wide Field of View Prototype for the SKA}

\author{S.A.~Torchinsky$^a$\thanks{Corresponding author}, A.O.H.~Olofsson$^b$, B.~Censier$^a$, A.~Karastergiou$^{c,d,e}$,
M.~Serylak$^{e,f}$, P.~Renaud$^a$, C.~Taffoureau$^a$\\
\llap{$^a$}Observatoire de Paris,
  Station de radioastronomie de Nan\c{c}ay, France\\
  \llap{$^b$}Onsala Space Observatory,
  Chalmers University of Technology, Sweden\\
  \llap{$^c$}Astrophysics, University of Oxford, Denys Wilkinson Building,
Keble Road, Oxford OX1 3RH, UK\\
\llap{$^d$}Department of Physics and Electronics, Rhodes University, PO Box
94, Grahamstown 6140, South Africa\\
\llap{$^e$}Department of Physics \& Astronomy, University of the Western Cape, Cape Town, South Africa\\
\llap{$^f$}SKA South Africa, Cape Town, South Africa\\
E-mail: \email{torchinsky@obs-nancay.fr}}

\abstract{
A revolution in radio receiving technology is underway with the
development of densely packed phased arrays for radio astronomy.  This
technology can provide an exceptionally large field of view, while at
the same time sampling the sky with high angular resolution.  Such an
instrument, with a field of view of over 100 square degrees, is ideal
for performing fast, all-sky, surveys, such as the ``intensity mapping''
experiment to measure the signature of Baryonic Acoustic Oscillations in
the HI mass distribution at cosmological redshifts.  The SKA, built
with this technology, will be able to do a billion galaxy survey.  I
will present a very brief introduction to radio interferometry, as
well as an overview of the Square Kilometre Array project.  This will
be followed by a description of the EMBRACE prototype and a discussion
of results and future plans.}


\begin{document}

\section{Introduction}

\noindent The Square Kilometre Array (SKA)~\cite{dewdneyska} will be
the largest radio astronomy facility ever built with more than 10
times the equivalent collecting area of currently available
facilities. The SKA will primarily be a survey instrument with
exquisite sensitivity and an extensive field of view providing an
unprecedented mapping speed. This capability will enormously advance
our understanding in fundamental physics including gravitation, the
formation of the first stars, the origin of magnetic fields, and it
will give us a new look at the Universe in the time domain with a
survey of transient phenomena.

\begin{figure}[ht!]
 \centering
 \includegraphics[width=0.95\linewidth,clip]{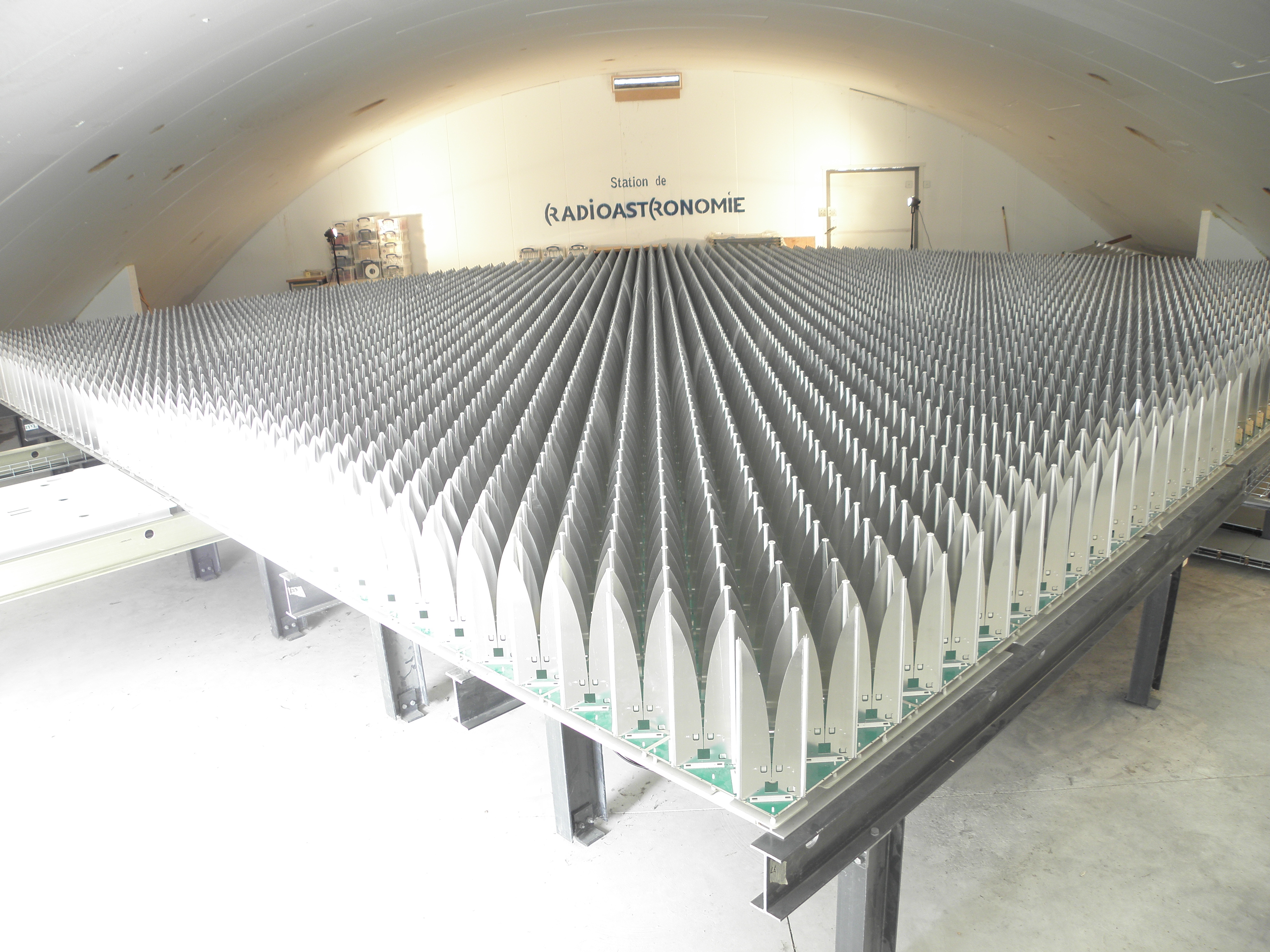}
  \caption{The EMBRACE array at \nancay\ is composed of 4608 Vivaldi antenna elements.}
  \label{fig:embrace}
\end{figure}

A revolution in radio receiving technology is underway with the
development of densely packed phased arrays.  This technology can
provide an exceptionally large field of view, while at the same time
sampling the sky with high angular resolution.  In addition to vastly
increased survey speed compared to traditional dish interferometers,
the aperture array technology provides the operational advantage
associated with a structure having no moving parts.  This is an
important consideration for the long term maintenance and operation of
a very large scale instrument like the SKA.  The \nancay\ radio
observatory is a major partner in the development of dense phased
arrays for radio astronomy, working closely with The Netherlands
Institute for Radio Astronomy (ASTRON).  The joint project is called
EMBRACE (Electronic MultiBeam Radio Astronomy Concept).  Two EMBRACE
prototypes have been built.  One at Westerbork in The Netherlands, and
one at \nancay\ (see Figure~\ref{fig:embrace}).  The EMBRACE prototypes are recognized as ``Pathfinders'' for the SKA project and both of them are
currently being extensively characterized and tested at the two sites.
Conclusions from the EMBRACE testing will directly feed into the SKA
and will have a decisive impact on whether or not dense array
technology is used for the SKA.

The date for selection of technology for the SKA is 2016.  If dense
arrays are not selected for the SKA, then the SKA will have a much
reduced mapping speed compared to what has come to be expected by the
astronomical community.  It is therefore of crucial strategic
importance that work on EMBRACE succeeds in showing the viability of
dense arrays for radio astronomy.

The two EMBRACE stations began with an initial period of engineering
testing on the partially complete arrays
\cite{olofsson_limelette,wijnholds_limelette}.
\enancay\ has been fully operational since 2011
and now performs regularly scheduled
astronomical observations such as pulsar observations and
extragalactic spectral line observing~\cite{2013sf2a.conf..439T}.
EMBRACE system characteristics such as beam main lobe and system
temperature are behaving as expected.  EMBRACE has long term
stability, and after four years of operation continues to prove itself
as a robust and reliable system capable of sophisticated radio
astronomy observations.

\section{Science Goals for SKA using EMBRACE Technology}
Aperture Arrays are the enabling technology which will lead to fulfilling the
promise of SKA as a transformational survey machine.  The Field of
View offered by an aperture array is well beyond that possible with
any other technology.  At the frequencies in the mid range of SKA, an
aperture array based on EMBRACE technology will have a Field of View
on the order of 100~square~degrees.  It is primarily the science
objectives which require large surveys of the sky which will benefit
most from the aperture array technology.

The speed at which the sky can be mapped depends on the sensitivity of
the instrument and also on the field of view.  It is possible to be
equally fast with a very sensitive telescope and a small field of
view, compared to a less sensitive telescope with a large field of
view.  The parameter known as the ``survey speed'' measures how fast a
telescope can survey the sky, but there is a basic assumption that the
sky does not change, and repointing the telescope to fill-in the map
can be done at any time.  This is not the case for a study of
transient phenomena.

\subsection{Dark Energy}
The original science driver for the SKA is the measurement of the mass
distribution at cosmological epochs by tracing the neutral hydrogen
gas (HI) at different redshifts~\cite{2004NewAR..48.1013R}.  Neutral
hydrogen is the most abundant material in the Universe, and a survey
of its distribution clearly gives information on the overall geometry
and evolution of the Universe.  However, the signal from HI is very
weak, and the radio sky is very noisy, and the only way to detect weak
signals in a noisy background is to have an extremely sensitive
telescope.  This fact was already recognized nearly fifty years ago by
Heidmann~\cite{1966LAstr..80..157H}.  Heidmann not only understood the
importance of detecting HI at cosmological distances, but he also
calculated that a radio telescope of nearly a square kilometre
collecting area was necessary to detect nearly a million extragalactic
sources.  This remains the primary objective of the Square Kilometre Array.

With the discovery that the Universe is expanding at an accelerating
rate~\cite{1998AJ....116.1009R}, the quest to understand Dark Energy became
preeminent in the physics community.  The effect of Dark Energy on the
geometry of the Universe can be traced by detections of the signature
of Baryonic Acoustic Oscillations (BAO) in measurements of the galaxy
distribution at different cosmological
epochs~\cite{2003ApJ...594..665B}.
The galaxy distribution can be measured using radio emission at
the wavelength of 21~cm produced during hyperfine transition of neutral
hydrogen,
and this has
the advantage of measuring both the position and redshift of galaxies
in a single observation, making the SKA the most reliable instrument
to perform the so-called ``wiggles'' experiment~\cite{2004NewAR..48.1013R}.

A galaxy survey in the redshift range between z=0.5 to z=3 is optimum
for the BAO experiment.  This corresponds to frequencies in the range
350~MHz to 1000~MHz, and continues naturally from the optimum high
frequency of the SKA Low Frequency Aperture Array.  

Moreover, a field of view on the order of 100's of square degrees, and
high angular resolution given by the $\sim200$~km baseline, makes the SKA
mid frequency aperture array the only instrument capable of performing this experiment after
only a few years observing.  The SKA catalog of over a billion
galaxies, with both their angular position and redshift, will be the
ultimate database for BAO analysis.  This will give exquisite
constraints on the Equation of State of the Universe, and may well
indicate that Einstein's Theory of General Relativity, together with a
Cosmological Constant, is insufficient to explain Dark Energy.

An intermediate experiment can be performed to make a statistical
detection of the BAO signature using the ``Intensity Mapping'' technique
\cite{2006astro.ph..6104P,2008arXiv0807.3614A}.  Instead of detecting
individual galaxies, and creating a catalog from which to perform the
analysis, it is sufficient to map the distribution of HI at different
epochs (i.e. different frequency bins).  The method is analogous to
the observational method used in Cosmology on measurements of the
anisotropy of the Cosmic Microwave Background.  The gas content in the
volume covered by a telescope beam in a given frequency band
(i.e. redshift bin) is correlated with measurements across the entire
sky giving a statistical detection of the BAO signature in the power
spectrum.  In this case, an equivalent collecting area of several
thousand square metres gives sufficient sensitivity to perform the
experiment provided the sky can be mapped rapidly (i.e. multibeam
observations, see~\cite{2012A&A...540A.129A}).  A beamsize on the order of a
few arcminutes is sufficent angular resolution, and indeed, to measure
the first peak in the BAO signature, a beamsize of 1 degree is all
that is necessary.

An precursor to the mid frequency aperture array would be ideally suited to do the HI Intensity
Mapping experiment.  Such an instrument would have a collecting area
of several thousand square metres, possibly organized in stations, and
confined to a core within a diameter of a few hundred metres.  The
mid frequency aperture array precursor, operational in the same timeline as SKA Phase 1,
could potentially deliver the first detection of BAO, directly
detected in HI which is the primordial material of the Universe.

\subsection{Pulsars and Transients}

Pulsar science can be broadly categorized into two areas: searches for
new pulsars and studies of known pulsars. Searches rely on two things,
raw sensitivity and sky coverage, which a dense aperture array can
provide as described already. Studies of known pulsars, also commonly
referred to as timing, rely on additional properties of the array,
such as timing/clock stability and polarization purity.  These are
essential parameters for using pulsar observations as tests of
theories of gravitation~\cite{2004NewAR..48.1459C}.  Clock stability has
already been demonstrated on the LOFAR high-band antennas.  Polarization purity
is a topic of active study, to ensure that observations off the dipole
axis are still calibratable to the extent required for pulsar timing.

The main advantage of a dense aperture array comes in its ability to
form large numbers of tied array beams across the sky. This would
allow a very efficient pulsar timing programme of newly discovered
millisecond pulsars, alleviating the issues faced by dish arrays that
can only simultaneously form tied-array beams within the primary beam
of a dish (see for example~\cite{2009A&A...493.1161S}).

The sky is filled with sources which are periodic, such as pulsars, or
are one-time catastrophic events such as merger events, accretion
events, supernovae, (see
eg.~\cite{2004NewAR..48.1459C,2015arXiv150104716C}) or the enigmatic
``Fast Radio Bursts'' first discovered by Lorimer et
al.~\cite{2007Sci...318..777L} (see also~\cite{2015arXiv150107535M}).
In order to observe such an event, the telescope must be pointing in
the right direction at the right time.  It's not possible to move the
telescope around to ``fill-in'' the sky.  The only way to maximize the
likelihood of detecting a transient event is to point as long as
possible in as many directions as possible, or in other words, to have
a very large field of view.  A large, dense aperture array using
EMBRACE type technology is therefore the ideal instrument to make a
survey of transient events.

\subsection{The Unknown}

Opening up phase space leads to new discoveries~\cite{2004NewAR..48.1551W}
and in fact, most telescopes have made unexpected
discoveries completely outside the domain for which they were
designed~\cite{torchinsky_limelette}.

The ``exploration of the unknown'' is more than just an exploration of
the time domain.  The aperture array technology provides an improvement all across
parameter-space: spectral coverage, field of view, time domain.  There
may well be surprises not only in time varying sources, but also in
unexplored parts of the spectrum, and in correlation across the sky.
An example from history is the discovery of the cosmic microwave
background.  It was not identified until it was measured across the
sky, even though we can see now that there were in fact measurements
of the CMB as early as 1940~\cite{1940PASP...52..187M}, well before the
famous detection by Penzias \& Wilson in
1965~\cite{1965ApJ...142..419P}, but pointed measurements were not enough to make people see that
something important had been measured.  It had to be seen at many,
widely separated, positions on the sky.  The wide field of view of aperture arrays may well
come up with similar surprises in the future, which are not
necessarily time domain related.

\section{A Very Brief Reminder of Interferometry}

\begin{figure}[ht!]
 \centering \includegraphics[width=0.45\linewidth,clip]{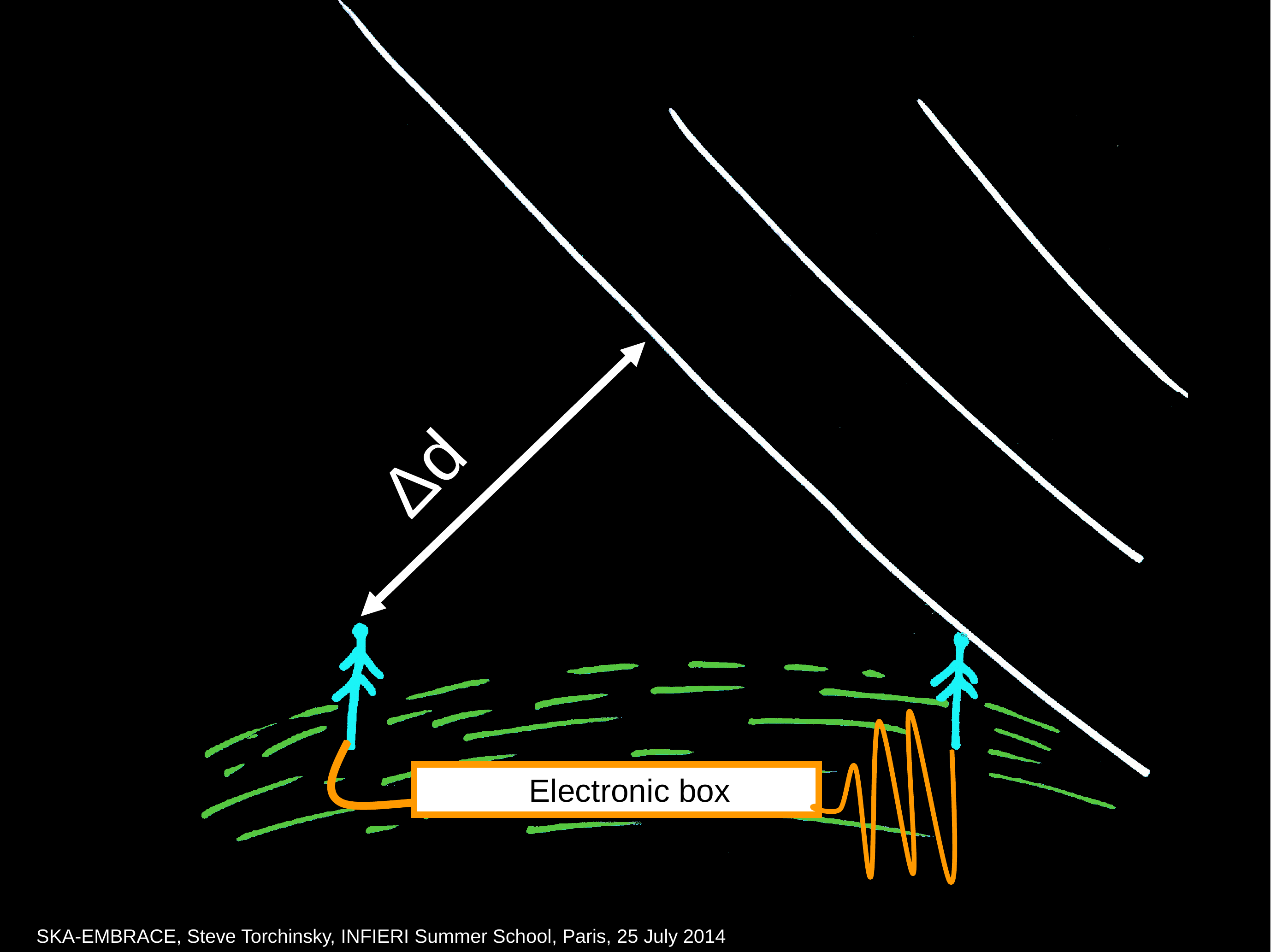} \includegraphics[width=0.45\linewidth,clip]{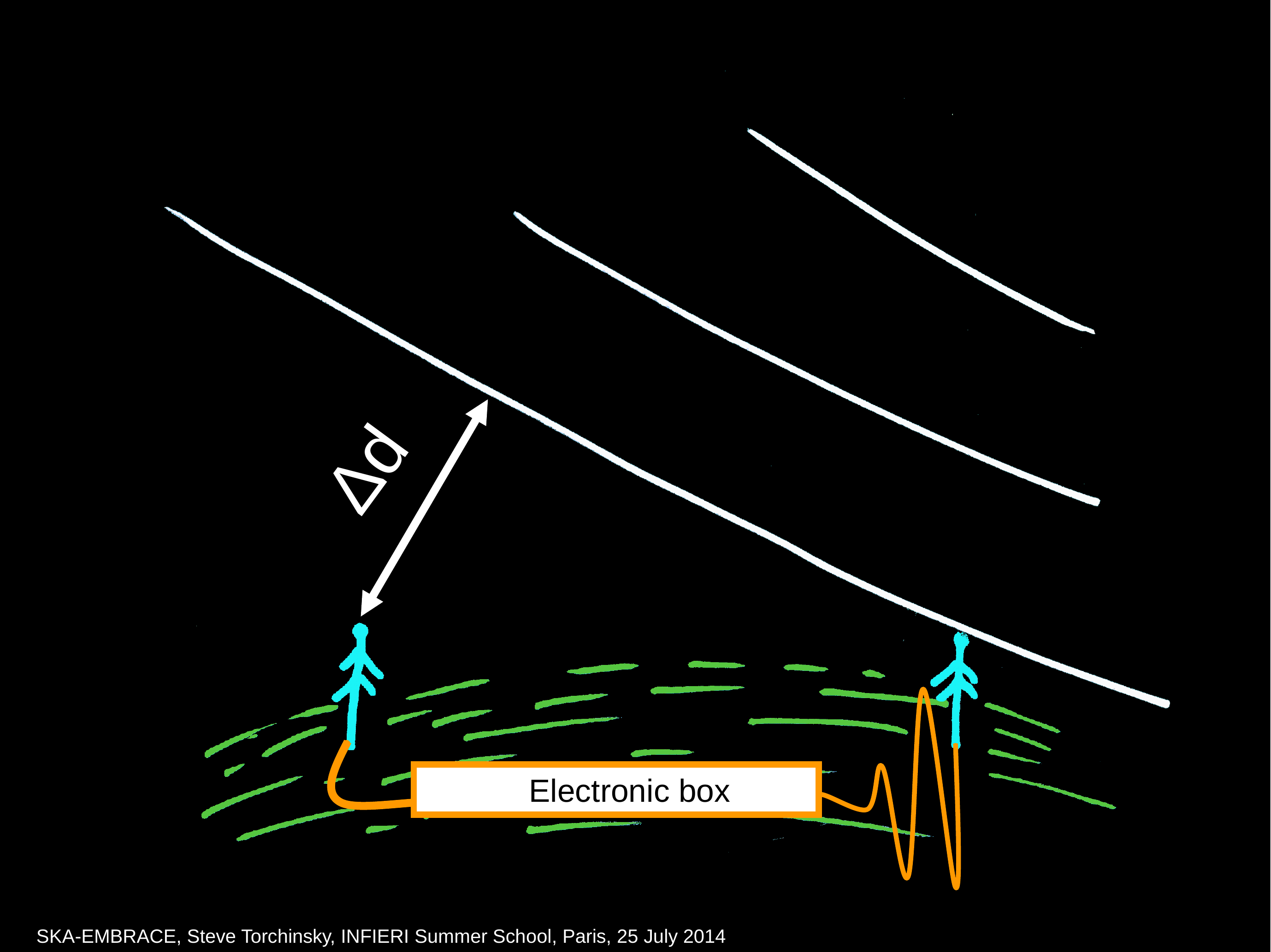} \caption{A
 plane wave from a distant source arrives at the Earth and the
 wavefront is first detected by one antenna, and then by the next one.
 The difference in the path length between the two receivers is
 dependent on the direction of arrival of the incident
 wavefront, and can be compensated by a length of cable before correlation.}
 \label{fig:delayline}
\end{figure}

The basic principle of interferometry is the combination of signals
from different antennas in such a way that the result is effectively
an instrument with a total collecting area equivalent to the combined
areas of all the antennas involved, and pointed in the desired
direction.  Figure~\ref{fig:delayline} illustrates how this can be
done in the simplest case of two antennas.  A plane wave from a
distant source arrives at the Earth and the wavefront is first
detected by one antenna, and then by the next one.  The difference in
the path length between the two receivers is dependent on the
direction of arrival of the incident wavefront.  The path length
difference can be compensated in electronics simply by adding a length
of cable corresponding to the desired path length difference.  By
switching in and out cables of different lengths, the array of
antennas are effectively pointing in different directions.  This is called ``beamforming''.

\begin{figure}[ht!]
 \centering
 \includegraphics[width=0.6\linewidth,clip]{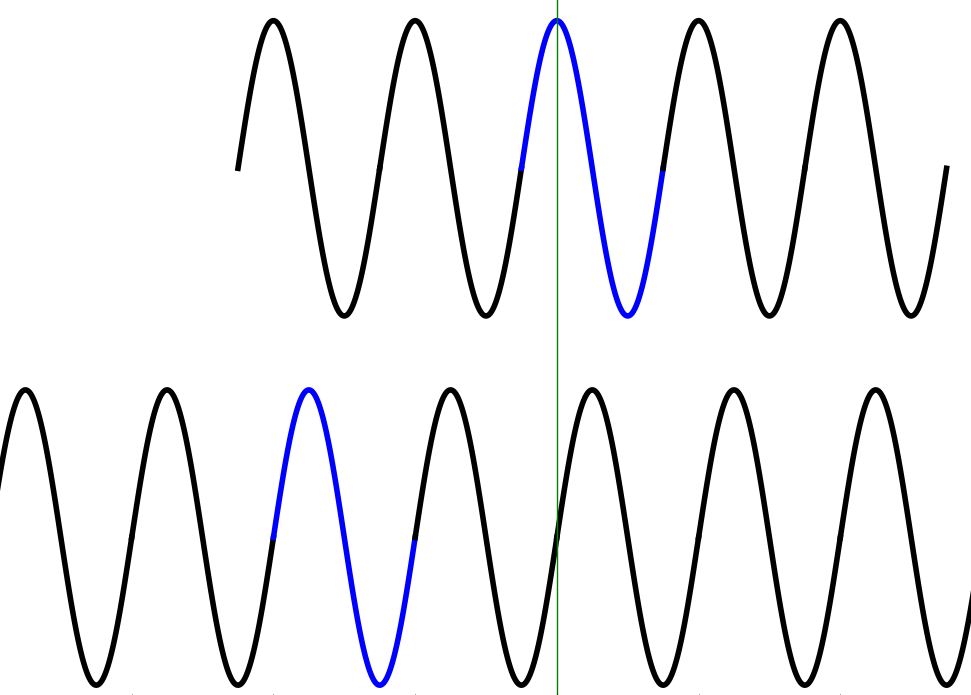}
  \caption{These two signals are out of phase as seen by peaks and troughs which do not line up.}
  \label{fig:phaseshift1}
\end{figure}

\begin{figure}[ht!]
 \centering
 \includegraphics[width=0.6\linewidth,clip]{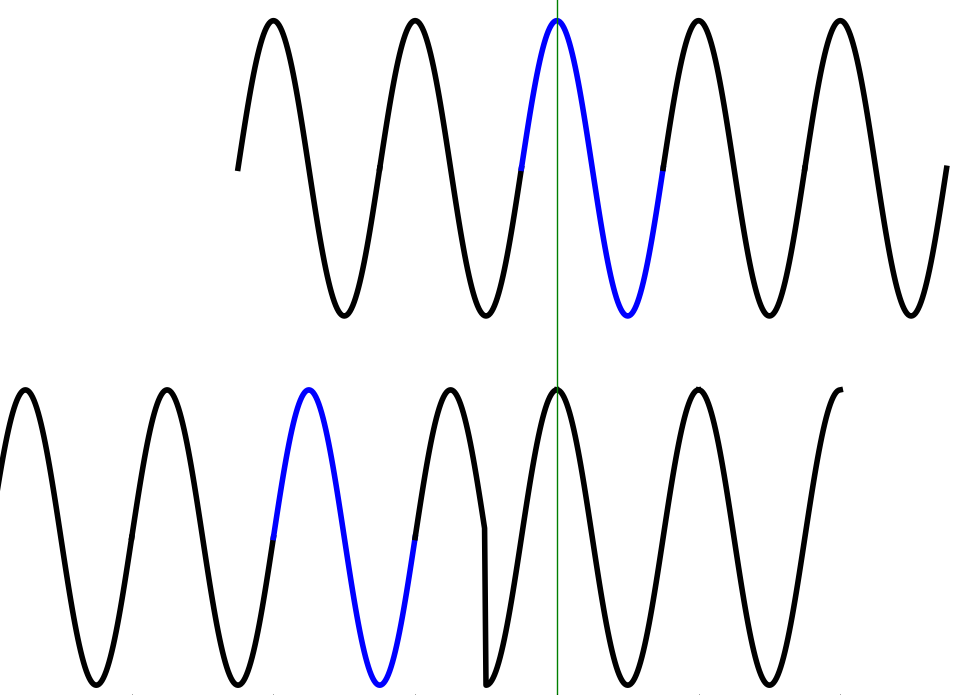}
  \caption{It is possible to create a phase shift on the second signal such that the peaks and troughs of the two signals line-up.  Note however that the signals are combined but not at the same wavefront (shown in blue).}
  \label{fig:phaseshift2}
\end{figure}

Instead of switching in and out cables, it is possible to make an
electronic circuit which makes a phase shift on the signal.
Figure~\ref{fig:phaseshift1} shows two signals which are not in phase.
Figure~\ref{fig:phaseshift2} shows a phase shift applied to the second
signal such that the signals are in phase and can be combined.  We
note however that the signals are combined but not at the same
wavefront (shown in blue).  This is permitted as long as the
separation between the wavefronts is within the ``coherence length''
which is defined as the inverse of the bandpass multiplied by the
speed of light.  For EMBRACE with frequency channels of 195~kHz, the
coherence length is on the order of 1.5~km, so using phase shifting
instead of a delay line is permitted.


For a detailed introduction and explanation of radio interferometry
and aperture synthesis, the reader is strongly recommended to refer
to the book by Thompson, Moran, and Swenson~\cite{interfero_bible}.

\section{EMBRACE System Description}

\enancay\ is a phased-array of 4608 densely packed antenna elements
(64 tiles of 72 elements each).  For mechanical, and electromagnetic
performance reasons, \enancay\ has, in fact, 9216 antenna elements,
but only one polarization (4608 elements) have fully populated signal
chains.  The orientation of the linear polarized elements at
\nancay\ is with the electric field sensitivity in the North-South
direction.  For more details on EMBRACE architecture see
\cite{kant_limelette, kant11}

\subsection{Hierarchical Analogue Beam Forming}

\enancay\ uses a hierarchy of four levels of analogue beamforming
leading to 16~inputs to the LOFAR~\cite{2013A&A...556A...2V} backend
system used for digital beamforming.

The first level of beamforming is done for four Vivaldi elements within the integrated
circuit ``beamformer chip'' developed at \nancay~\cite{bosse10}.
This chip applies the phase shifts necessary to four antenna elements
to achieve pointing in the desired direction.  The phase shift
required for each Vivaldi element is calculated from the array
geometry of a tile for a given pointing direction
\cite{pezzani_beamforming}.

The beamformer chip forms two independent beams for each set of four
antenna elements.  The beamformer chip splits the analogue signal from
the antennas into two signals and then applies different sets of phase
shift parameters to each signal.  The beamformer chip therefore has
four inputs for four Vivaldi antennas, and two independent outputs.
The independent outputs is what makes it possible for EMBRACE to
have two independent Fields~of~View, often referred to as
``RF~Beams'', which are named Beam-A and Beam-B.

The output of 3 beamformer chips is summed together on a ``hexboard''
and 6 hexboards make a tile. The EMBRACE array at \nancay\ has a
further analog summing stage with 4~tiles making a tileset.  This
final stage is done on the Control and Down Conversion (CDC) card in
the shielded container which is connected to the tiles via 25~m long
coaxial cables.

\subsection{Control and Down Conversion}
\label{sec:CDC}
The CDC cards are responsible for three
important tasks~\cite{bianchi_limelette, monari_limelette}:
1)~Frequency mixing the Radio Frequency (RF) for conversion to a
100~MHz bandwidth centred at 150~MHz; 2)~48~Volt Power distribution to
the tiles; 3)~Distribution of command and housekeeping data
communication to the tiles.  The RF, 48~Volt power supply, and
ethernet protocol monitoring and control communication are all
multiplexed on the coaxial cables connecting the tiles to the CDC
cards~\cite{berenz_limelette}.

\subsection{Digital Processing}
\label{sec:digproc}
The output of the tilesets is fed into a LOFAR-type digital Receiver
Unit (RCU) and Remote Station Processing (RSP) system for
digital beamforming~\cite{picard_limelette}.  The RSP performs the
digital beamforming of the entire array, producing pencil beams which
are usually called ``digital beams'' in LOFAR parlance.  These digital
beams can have pointings within the beam produced by the individual
inputs to the RCU.  For \enancay, the inputs to each RCU is the signal
from a tileset with a beam width of approximately 8.5\degrees.

EMBRACE is a single polarization instrument but the LOFAR RSP system
has the capacity to produce two outputs per digital beam which
correspond to the two orthogonal linear polarizations in LOFAR.  These are called
the ``X'' and ``Y'' beams.  For EMBRACE, ``X'' and ``Y'' are two,
possibly different, pointing directions within the field of view (the
RF~beam).

Fast data acquisition from the RSP boards is done by the backend
called the Advanced Radio Transient Event Monitor and Identification
System (ARTEMIS) developed at Oxford University
\cite{2013IAUS..291..492S}.

The ARTEMIS hardware is also used for recording raw data packets from
the RSP boards giving the digitized beam formed wavefront data.  This
is used for the high spectral resolution observation of the extra
galactic source M33.

\subsubsection{Statistics Data}
\label{sec:statsdata}
In addition to the high rate beamformed data produced by the RSP system,
there are also slower cadence data produced at a rate of once per second:
the crosslet statistics, the beamlet statistics, and the subband statistics.

The LOFAR RCU system digitizes and channelizes the 100~MHz wide RF bandpass
into 512 so-called subbands, each of 195.3125~kHz bandwidth. The cross
correlations of all tilesets are calculated once per second and are called
crosslets. The default mode of operation for LOFAR is to calculate the
crosslets for each subband in succession such that it takes 512 seconds to
cycle through the full RF band.

Another possibility is to request a given subband, and the crosslets
are calculated each second for the same subband. This is the mode of
operation used most often at \enancay. The subband statistics is
simply the 100~MHz bandpass for each tileset. The beamlet statistics,
called beamlets, are the beamformed data for the full array (16~inputs
for \enancay) dumped at 1~second intervals, and are identical to
the fast data integrated over 1~second.

\subsection{EMBRACE Monitoring and Control Software}

The Monitoring and Control software for EMBRACE was developed at
\nancay.  An extensive Python package library on the SCU (Station
Control Unit) computer gives scripting functionality for users to
easily setup and run observations depending on e.g.
type of the target.  Integrated statistics data are acquired from the LCU
(Local Control Unit) and saved into FITS files including header
information with essential meta data including pointing information,
timestamp, frequencies, etc.  Raw data (beamlets) are captured from
LCU Ethernet 1~Gbps outputs and saved into binary files
\cite{renaud_MAC}.

\section{Observations with \enancay}

\subsection{Multibeams with \enancay}

\begin{figure}[ht!]
 \centering
 \includegraphics[width=0.9\linewidth,clip]{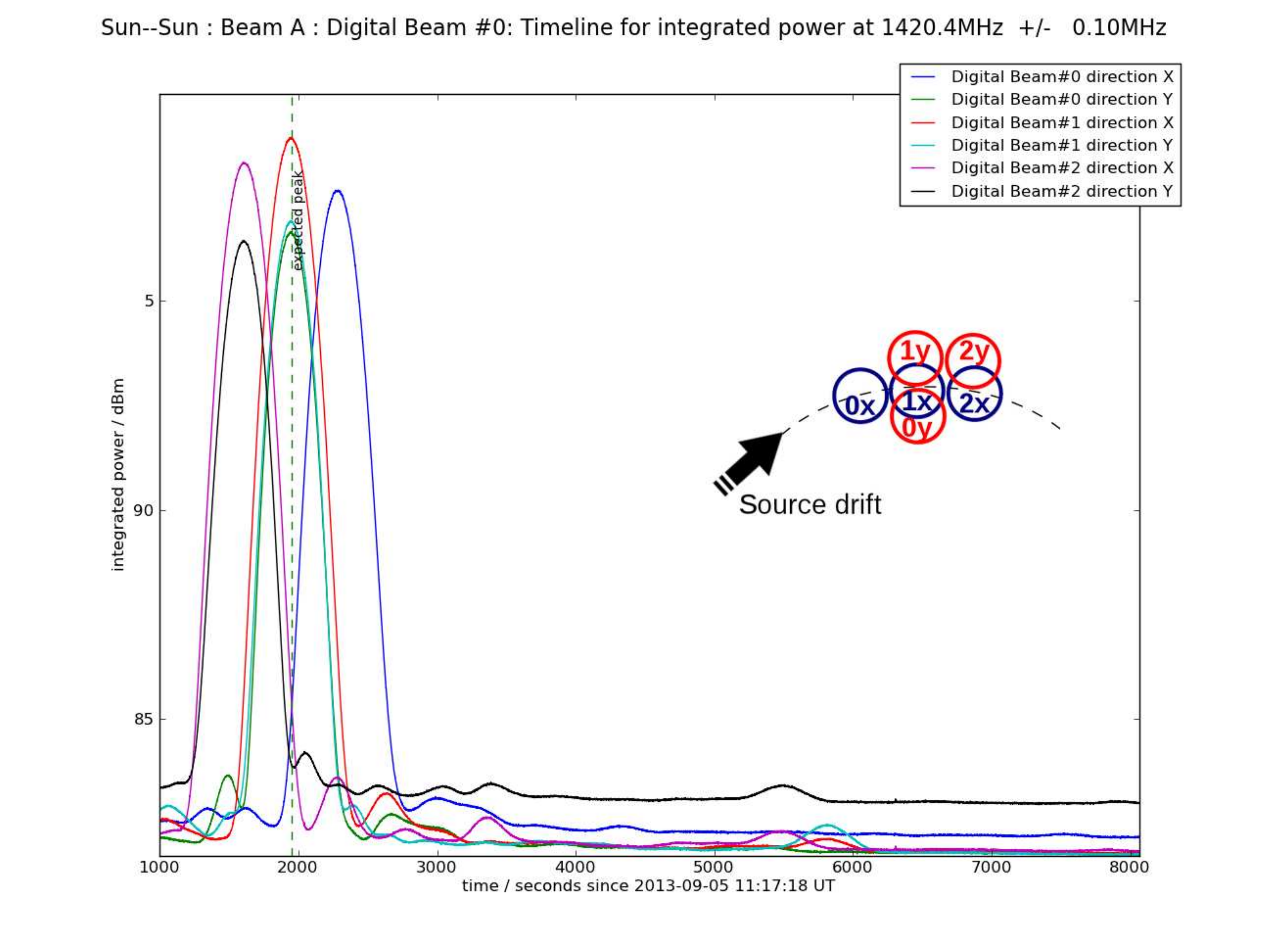}      
  \caption{This drift scan of the Sun used 6~beams of
    \enancay\ pointing along the trajectory of the Sun across the sky
    (inset top right).}
  \label{fig:multidrift}
\end{figure}

Figure~\ref{fig:multidrift} shows a drift scan of the Sun using the
multibeam capability of \enancay.  Six beams were pointed on the sky
along the trajectory of the Sun, including three partially overlapping
beams.  The result shows the Sun entering and exiting each beam and
the off-pointed beams are 3~dB down from the peak, as expected.

\begin{figure}
\centering
\includegraphics[width=0.6\linewidth,clip]{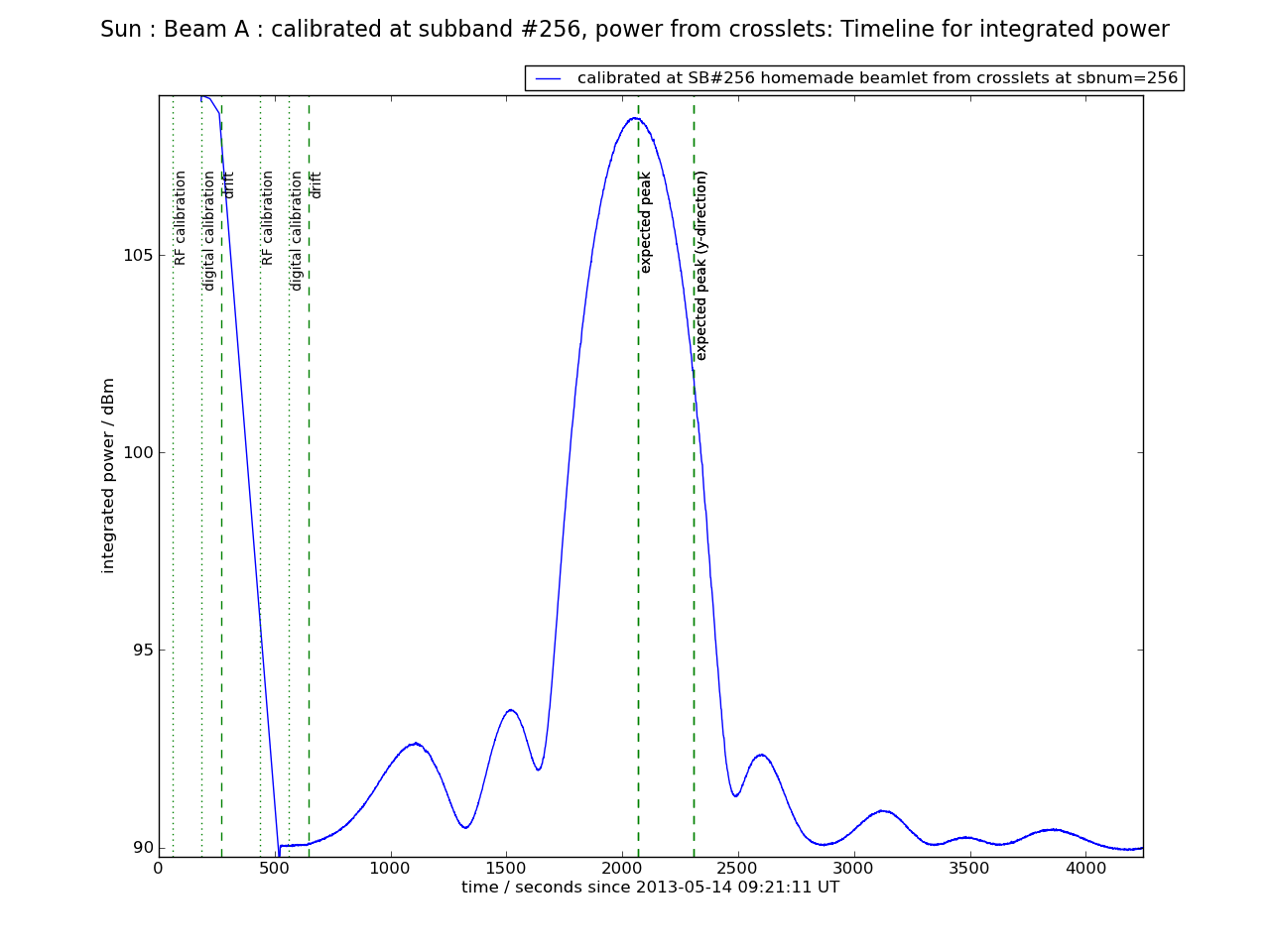}
\caption{
The drift scan of the Sun across the stationary digital beam pointing
reaches its maximum at the expected time.  The sidelobe levels are
below the 15~dB design criterion.
}
\label{fig:sundrift}
\end{figure}

\subsection{Pulsar B0329+54}
\label{sec:psr}
\begin{figure}[ht!]
 \centering \includegraphics[width=0.6\linewidth,clip]{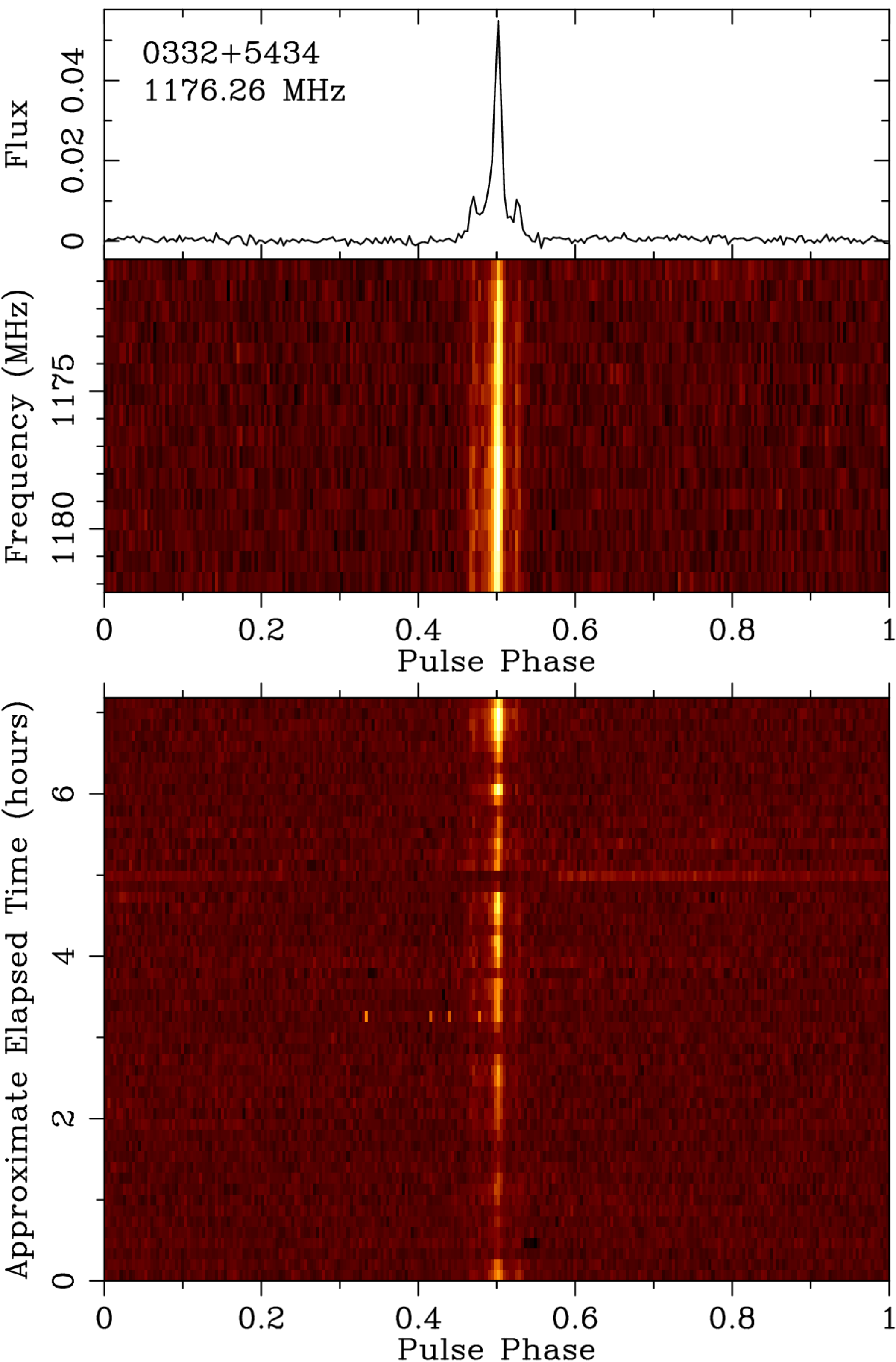}
\caption{Pulsar B0329+54 was detected after several minutes as shown in this dynamic
 plot folded at the pulsar period of 715~msec.  The figure was plotted
 using PSRCHIVE
 software~\cite{psrchive,psrchive2012}.}  \label{fig:b0329}
\end{figure}

Figure~\ref{fig:b0329} demonstrates over 9 hours of tracking
the pulsar B0329+54. Its pulsed signal is clearly detected after
several minutes, and the array continues tracking, measuring
continuously the pulsar, except where RFI has been filtered at 21500
seconds. The array was configured with a bandwidth of 12~MHz (62
beamlets) centred at 1176.45~MHz.

\begin{figure}[ht!]
 \centering
 \includegraphics[width=0.6\linewidth,clip]{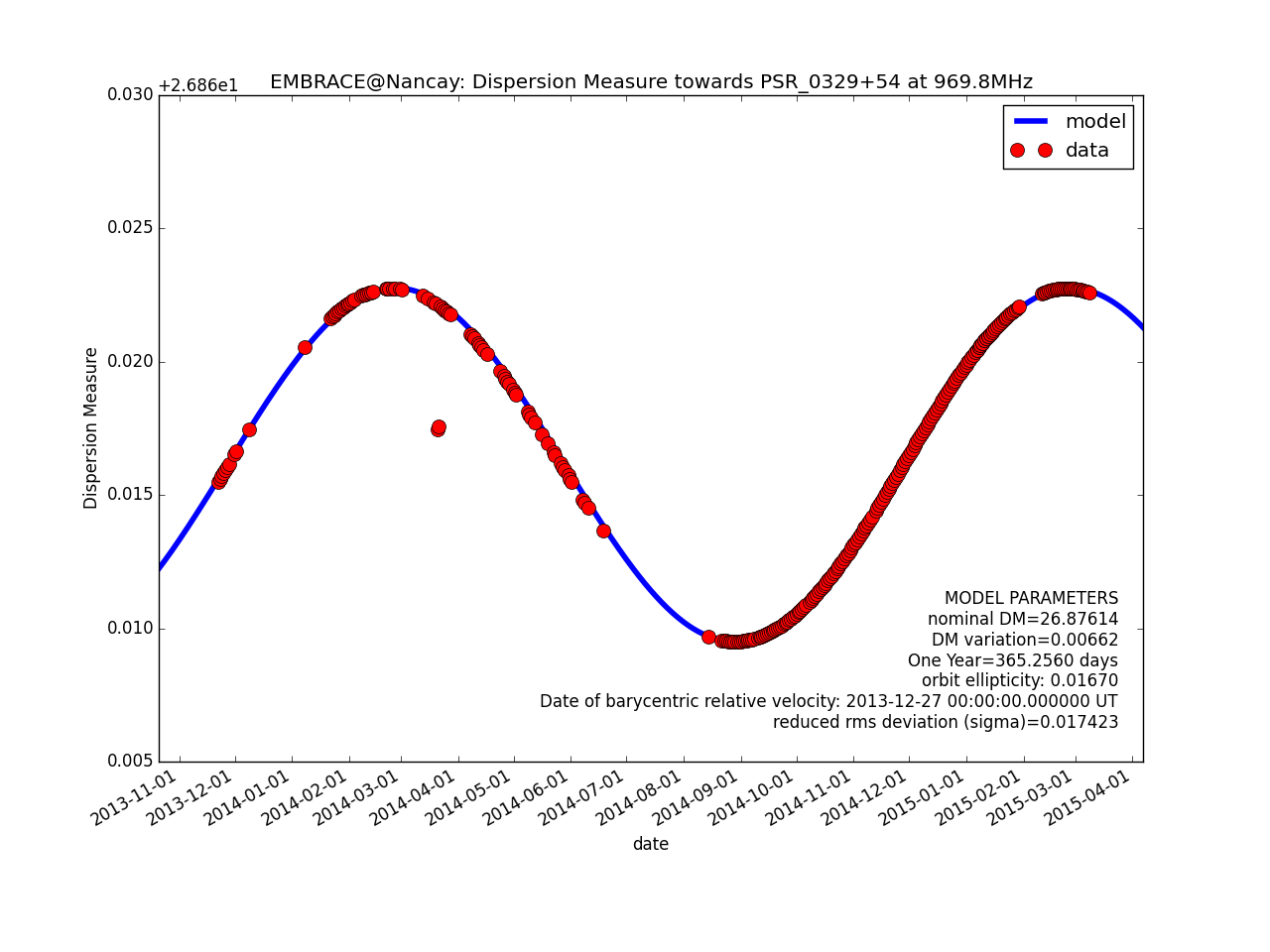}
  \caption{The Dispersion Measure towards Pulsar B0329+54 was measured each day since November 2013.  The annual variation of DM is consistent with what is expected by the Doppler Effect due to the movement of the Earth in orbit around the Sun.}
  \label{fig:dailyDM}
\end{figure}

In November 2013, \enancay\ began a long term campaign of observation
of pulsar B0329+54.  \enancay\ has been running autonomously, and
performing the pulsar observation every day.  Figure~\ref{fig:dailyDM}
shows the Dispersion Measure (DM) towards B0329+54 as determined
during each daily observation over the course of 16~months.  The
annual variation of DM is consistent with what is expected by the
Doppler Effect due to the movement of the Earth in orbit around the
Sun.

Radiation traveling through the Interstellar Medium (ISM) between a
source object, such as a pulsar, and the Earth, is subject to a
retarding effect due to the plasma in interstellar space.  The ISM
behaves as a refractive medium which is slowing the propagation of the
electromagnetic wave in a manner which is inversely dependent on the square of the frequency (see for
example~\cite{2002ASPC..278..251S}).

The movement of the Earth due to its orbit around the Sun introduces a
Doppler Effect which modifies the signal frequency measured at the
telescope compared to its value in the ISM.

The delay between reception of the high and low frequencies is also
subject to the Doppler effect due to the Earth's motion.  The Earth is ``catching up'' or ``moving away'' from the signal, and so the time between reception of the high and low frequencies is reduced or delayed.

The value of the Dispersion Measure varies throughout the year due to the Earth's
orbit around the Sun 

as a function of the projected velocity onto the line of site to the
pulsar.
Taking the Earth's elliptical orbit into consideration, and
the direction of pulsar B0329+54, gives a variation of Dispersion
Measure of 0.0066 on a nominal value of $26.8761pc/cm^{3}$.  The model and data points are plotted in Figure~\ref{fig:dailyDM}.  Dispersion Measure was calculated from the ARTEMIS filterbank files using the PRESTO suite of tools~\cite{ransom_phd}.

\subsection{The Triangulum Galaxy: Messier~33}
\begin{figure}
\centering
\includegraphics[width=\linewidth,clip]{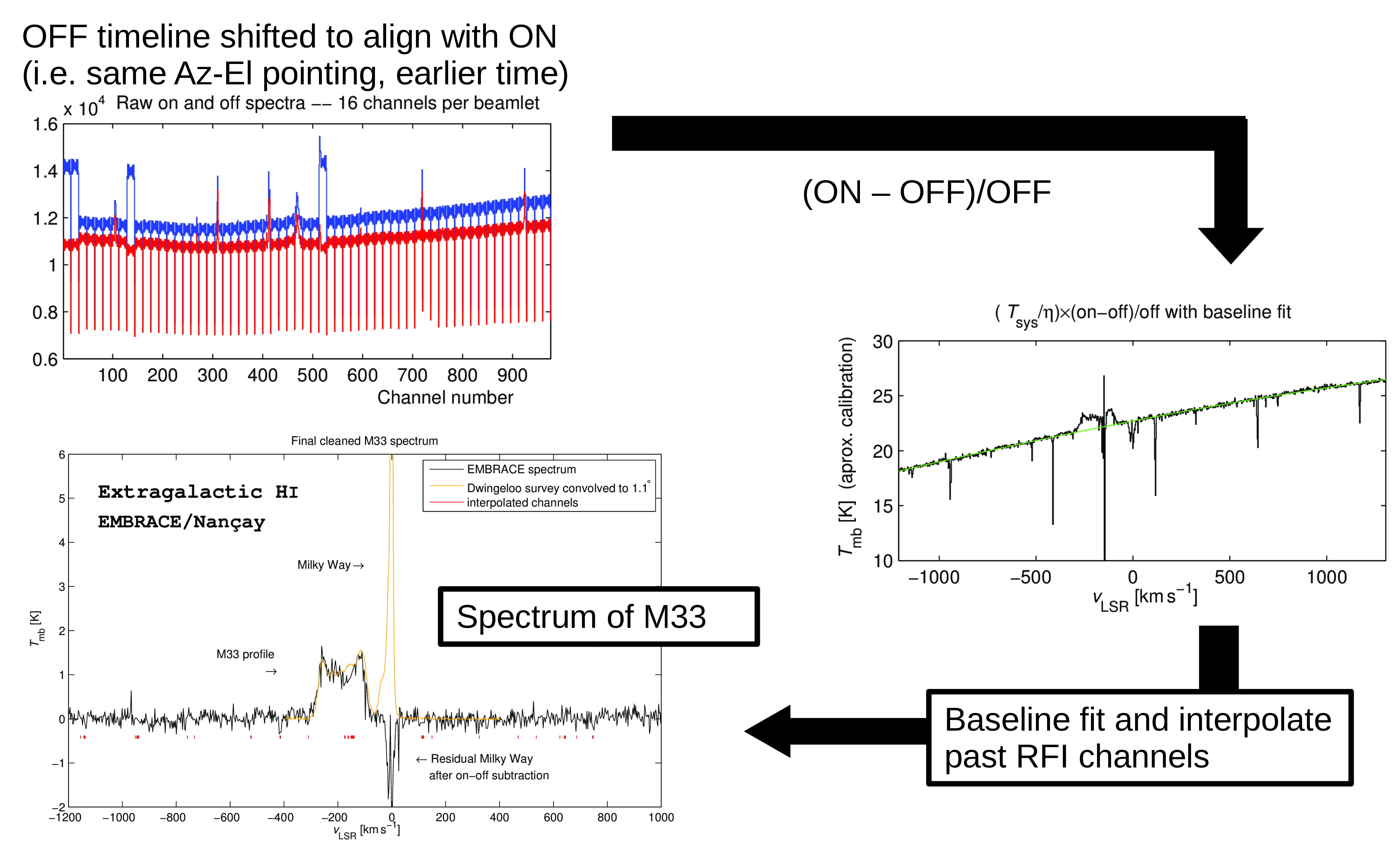}
\caption{\textbf{Top left:} Raw averaged spectra (blue=on, red=off).
Note that the horizontal axis has increasing frequency to the right. The
galactic foreground is visible in both positions around channel 470.
\textbf{Mid right:} Coarsely calibrated spectrum with four offset subbands
corrected by adjusting their level to that of neighbouring subbands.
Also shown in green is a second order baseline fit. \textbf{Bottom left:}
Baseline subtracted spectrum where obvious spikes have been interpolated out.
A comparison with the LAB survey spectrum is shown in golden/yellow.
The detailed line shape agreement is excellent apart from a portion that
is obviously asscociated with an unstable subband. Note that the galactic
foreground has mostly been subtracted out in the EMBRACE spectrum with only
a negative residual remaining.}
\label{fig:m33}
\end{figure}
\begin{figure}
\centering
\includegraphics[width=0.45\linewidth,clip]{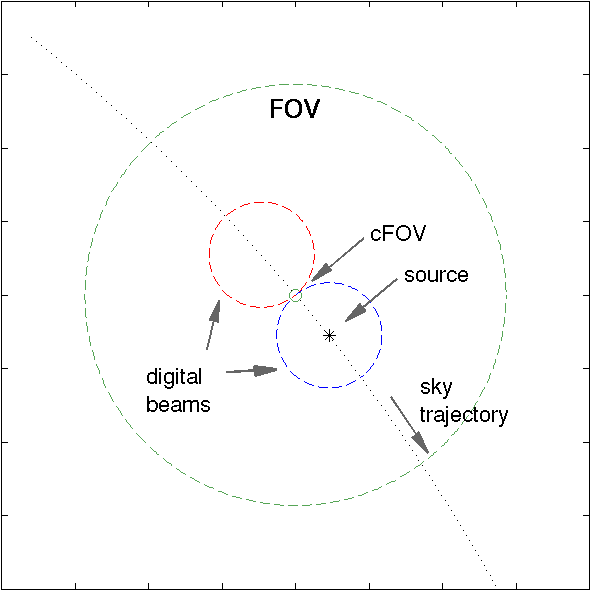}
\caption{Two digital beams are placed symmetrically within the FoV
(tileset beam) following the same sky trajectory but with a time delay.
One of them is centred on the source, the other is used as an off-position.}
\label{fig:obsstrat}
\end{figure}

During the summer of 2013, M33 was observed at high spectral resolution by capturing the stream of UDP packets for half an hour
(160~Gbyte in total) and channelizing each subband 16-fold (see Figure~\ref{fig:m33}).
The two digital beams that are always present in the backend output
(referred to as the ``X'' and ``Y'' directions due to its LOFAR heritage)
were employed to create symmetric on- and off-beams in a manner described
in Figure~\ref{fig:obsstrat}.
A spectrum was subsequently created in the simplest way possible; by
co-averaging all data for each position separately before off-position
subtraction and normalization, and then multiplying by a constant factor
so that our result matches a template spectrum we retrieved from the
LAB survey~\cite{2005A&A...440..775K} convolved to the
EMBRACE array beam size.

\section{Summary and Future Work}

The technology of dense aperture array uses a large number of antenna
elements at half wavelength spacings to fully sample the aperture.
\enancay\ is the first fully operational demonstrator of this
technology which is large enough to make interesting radio
astronomical detections.

One of the main concerns with the dense aperture array technology is
the high system complexity which could complicate operations and some
have expressed doubt that operating such a complicated instrument might not be
feasible for a facility observatory.  However, \enancay\ has clearly
shown that dense aperture array technology is perfectly viable for
radio astronomy.
We have demonstrated its capability as a radio astronomy instrument,
including astronomical observations of pulsars and spectroscopic
observations of galaxies.  We have also demonstrated its multibeam
capability.

An important disadvantage of the dense aperture array technology is
its relatively large power requirement.  The large number of analogue
electronic components associated with the signal chain of each antenna
element, together with the digital processing requirements for
beamforming and/or aperture synthesis, makes a system which is rather
power hungry.  A number of solutions are being studied to reduce the
overall power consumption~\cite{vaate2014}.

Dense aperture array technology is a viable solution for the SKA,
offering the benefit of an enormous field of view together with
great flexibility for system setups, including multibeaming with
multiple and independent observation modes.  The SKA built using dense
aperture array technology will be the most rapid astronomical survey
machine.

\acknowledgments
  EMBRACE was supported by the European Community Framework Programme
  6, Square Kilometre Array Design Studies (SKADS), contract no
  011938.  We are grateful to ASTRON for initiating and developing the
  EMBRACE architecture.  Maciej Serylak acknowledges the financial assistance of the South African
SKA Project (SKA SA).  Henrik Olofsson, Aris Karastergiou, and Maciej Serylak were supported for
  multiple working visits to \nancay\ by grants from the Scientific
  Council of the Paris Observatory.

\newcommand{\aap}{{\it A. \& A.}}
\newcommand{\apj}{{\it Ap.~J.}}
\newcommand{\pasa}{{\it Publ. Astron. Soc. Australia}}
\newcommand{\pasp}{{\it Proc. Astron. Soc. Pacific}}
\newcommand{\nar}{{\it New Astronomy Reviews}}
\newcommand{\aj}{{\it Astr. J.}}
\newcommand{\limelettebib}{
2010 in {\it Proc. Wide Field Science and Technology for the SKA},
Limelette, Belgium, 
S.A.~Torchinsky \etal\ (eds),
ISBN~978-90-805434-5-4
}


\begin{thebibliography}{9}

\bibitem{dewdneyska}Dewdney,~P.E., Hall,~P.J., Schilizzi,~R.T., Lazio,~T.J.W.,
 {\em Proceedings of the IEEE}, Vol.~97,  No.~8,  pp.~1482-1496, ISSN:~0018-9219

\bibitem{olofsson_limelette}
Olofsson,~A.O.H. \TorchiSA, Chemin,~L. Barth,~S. Bosse,~S., Martin,~J.-M., Paule,~W., Picard,~P., Pomar\`ede,~S., Renaud,~P., Taffoureau,~C., Kant,~G.W., Noordam,~J.E., Wijnholds,~S.J., Keller,~R., Montebugnoli,~S.
\limelettebib, pg. 253

\bibitem{wijnholds_limelette}
Wijnholds,~S.J., Kant,~G.W., van~der~Wal,~E., Benthem,~P., Ruiter,~M., Picard,~P., Torchinsky,~S.A., Montebugnoli,~S., Keller,~R.,  \emph{EMBRACE: First Experimental Results with the Initial 10\% of a 10,000 Element Phased Array Radio Telescope,} \limelettebib, pg.~259

\bibitem{2013sf2a.conf..439T}
  Torchinsky, S.~A., Olofsson, A.~O.~H., Karastergiou, A., et al.\ 2013, SF2A-2013: Proceedings 
of the Annual meeting of the French Society of Astronomy and Astrophysics, 
439 

\bibitem{2004NewAR..48.1013R}
  Rawlings, S., Abdalla, F.~B., Bridle, S.~L., et al.\ 2004, \nar, 48, 1013 

\bibitem{1966LAstr..80..157H}
  Heidmann, J.\ 1966, L'Astronomie, 80, 157 

\bibitem{1998AJ....116.1009R}
  Riess, A.~G., Filippenko, A.~V., Challis, P., et al.\ 1998, \aj, 116, 1009 

\bibitem{2003ApJ...594..665B} Blake, C., \& Glazebrook, K.\ 2003, \apj, 594, 665 

\bibitem{2006astro.ph..6104P}
  Peterson, J.~B., Bandura, K., \& Pen, U.~L.\ 2006, arXiv:astro-ph/0606104 

\bibitem{2008arXiv0807.3614A}
  Ansari, R., Le Goff, J.~-., Magneville, C., et al.\ 2008, arXiv:0807.3614 

\bibitem{2012A&A...540A.129A} Ansari, R., Campagne, J.~E., Colom, P., et al.\ 2012, \aap, 540, AA129 
  
\bibitem{2004NewAR..48.1459C}
  Cordes, J.~M., Lazio, T.~J.~W., \& McLaughlin, M.~A.\ 2004, \nar, 48, 1459 

\bibitem{2009A&A...493.1161S}
  Smits, R., Kramer, M., Stappers, B., et al.\ 2009, \aap, 493, 1161 

\bibitem{2015arXiv150104716C} Corbel, S., 
Miller-Jones, J.~C.~A., Fender, R.~P., et al.\ 2015, arXiv:1501.04716 

\bibitem{2007Sci...318..777L}
  Lorimer, D.~R., Bailes, M., McLaughlin, M.~A., Narkevic, D.~J., \& Crawford, F.\ 2007, Science, 318, 777 

\bibitem{2015arXiv150107535M}
  Macquart, J.-P., Keane, E., Grainge, K., et al.\ 2015, arXiv:1501.07535 

\bibitem{2004NewAR..48.1551W} Wilkinson, P.~N., 
Kellermann, K.~I., Ekers, R.~D., Cordes, J.~M., 
\& W.~Lazio, T.~J.\ 2004, \nar, 48, 1551 

\bibitem{torchinsky_limelette}
  \TorchiSA, \emph{The Questions that Drive the Specifications} \limelettebib, pg.~33

\bibitem{1940PASP...52..187M} McKellar, A.\ 1940, \pasp, 52, 187 

\bibitem{1965ApJ...142..419P} Penzias, A.~A., \& Wilson, R.~W.\ 1965, \apj, 142, 419 

\bibitem{interfero_bible}
  Thompson,~A.R., Moran,~J.M., Swenson,~G.W.Jr., \emph{Interferometry and Synthesis in Radio Astronomy, 2nd Edition}
  Wiley-VCH, ISBN: 978-0-471-25492-8

\bibitem{kant_limelette}
Kant,~G.W., van~der~Wal,~E., Ruiter,~M.,  Benthem,~P. 2010
\limelettebib, pg.~227

\bibitem{kant11}
Kant,~G.W., Patel,~P.D., Wijnholds,~S.J., Ruiter,~M., van~der~Wal,~E. 2011
{\it IEEE Trans. A\&P} {\bf 59}, 1990.

\bibitem{2013A&A...556A...2V}
  van Haarlem, M.~P., Wise, M.~W., Gunst, A.~W., et al.\ 2013, \aap, 556, AA2 

\bibitem{bosse10} Bosse,~S., Barth,~S., Torchinsky,~S.A., Da~Silva,~B. {\emph Proc. European Microwave Integrated Circuits Conference,} (EuMIC 2010), 27-28 September 2010, Paris, France, pp.~106~-~109 

\bibitem{pezzani_beamforming}
Pezzani,~J., \emph{Algorithm for EMBRACE beamforming} EMBRACE Technical Document, \nancay, 2008

\bibitem{bianchi_limelette} 
Bianchi,~G., Morawietz,~J., Mariotti,~S., Perini~F., Schiaffino,~M., Kant,~~G.W.,  \emph{EMBRACE Local Oscillator distributor,} \limelettebib, pg.~249

\bibitem{monari_limelette} Monari,~J., Perini,~F.,
  Mariotti,~S., Kant,~G.W., Morawietz,~J., van~der~Wal,~E.  ``EMBRACE
  receiver design,'' \limelettebib, pg.~245

\bibitem{berenz_limelette} 
Berenz,~T., \emph{Mixed signal transportation for the EMBRACE antenna tiles} \limelettebib, pg.~239

\bibitem{picard_limelette}
Picard,~P., Renaud,~P., Taffoureau,~C., Macaire,~V., Mercier,~L., Paule,~W.
\limelettebib, pg. 235

\bibitem{2013IAUS..291..492S}
Serylak, M., Karastergiou, A., Williams, C., et al.\ 2013, IAU Symposium, 291, 492 

\bibitem{renaud_MAC}
Renaud,~P. Taffoureau,~C., Picard,~P., Borsenberger,~J., \TorchiSA, Olofsson,~A.O.H., Viallefond,~F., \emph{Monitoring and Control of EMBRACE, a 4608 Elements Phased Array for Radio Astronomy,} 2011, Proceedings Astronomical Data Analysis Software Systems, Paris, 6-10 November

\bibitem{psrchive}
  Hotan,~A.~W., van~Straten,~W., \& Manchester,~R.~N.\ 2004, \pasa, 21, 302 

\bibitem{psrchive2012}
  van~Straten, W., Demorest, P., \& Oslowski, S.\ 2012, Astronomical Research and Technology, 9, 237 

\bibitem{2002ASPC..278..251S} Stairs, I.~H.\ 2002, 
Single-Dish Radio Astronomy: Techniques and Applications, 278, 251 

\bibitem{ransom_phd}
  Ransom, S.~M.\ 2001, Ph.D.~Thesis,  Harvard University, ISBN: 9780493408415 

\bibitem{2005A&A...440..775K}
Kalberla, P.~M.~W., Burton, W.~B., Hartmann, D., et al.\ 2005, \aap, 440, 775 

\bibitem{vaate2014}
  bij~de~Vaate,~J.G., \TorchiSA, Faulkner~A.J, Zhang,~Y. Gunst,~A., Benthem,~P., van~Bemmel,~I.M., Kenfack,~G.
\emph{SKA Mid Frequency Aperture Arrays: Technology for the Ultimate Survey Machine}
in Proc. URSI General Assembly Beijing, China, 16-23 August, 2014 

  
\end{thebibliography}
\end{document}